# A Survey of Service Oriented Architecture Systems Testing

Ebrahim Shamsoddin-Motlagh[1]

[1]Computer Engineering Department, Faculty of Engineering, Science & Research Branch, Islamic Azad University, Tehran, Iran
`e.shamsoddin@srbiau.ac.ir`

## ABSTRACT

*Service oriented architecture (SOA) is one of the latest software architectures. This architecture is created in direction of the business requirements and removed the gap between softwares and businesses. The software testing is the rising cost of activities in development software. SOA has different specifications and features proportion of the other software architectures. First this paper reviews SOA testing challenges and existing solution(s) for those challenges. Then that reports a survey of recent research to SOA systems testing, that covers both functional and non-functional testing. Those are presented for different levels of functional testing, including unit, integration, and regression testing.*

## KEYWORDS

*Service Oriented Architecture Systems Testing, SOA*

## 1. INTRODUCTION

Arasanjani, Borges and Holley define SOA as follows [1]: "SOA is the architectural style that supports loosely coupled services to enable business flexibility in an interoperable, technology-agnostic manner. SOA consists of a composite set of business-aligned services that support a flexible and dynamically re-configurable end-to-end business processes realization using interface-based service descriptions."

System services have features in the design and implementation, these features include: Service reusability, Standardized service contract, Service loose coupling, Service abstraction, Service composition, Service autonomy, Service statelessness and service discoverability capabilities.
The purpose of this paper is to find the testing challenges in the SOA systems and existing solution(s) and to review recent research in the test of SOA systems. The paper is structured as follows. Section 2 challenges are expressed in the SOA Testing, and then the solutions to eliminate or minimize problems are expressed in these challenges. Section 3 a review related work of SOA Testing. Finally, Section 4 summarizes the paper and outline suggests future research steps.

## 2. SOA SYSTEMS TESTING CHALLENGES

The SOA system has different nature and the specific characteristics than the traditional system of the test system; it's harder and needs more time. The test facilitates and abilities at the SOA system testing should be recognized and solution(s) should be presented for testing challenges.

The key issues of testability limits of the SOA systems include: dynamicity and adaptiveness, lack of observability of the service code and structure, lack of control, lack of trust, new aspects of testing, test cost, different stakeholders [2, 3, 4, 5, 6, 7].

## 2.1. Functional Testing Challenges

The challenges of functional testing can be reported: All inputs and output may be very difficult to the system testing, Asynchronous in the SOA systems, rapid growth and large system (the development and management should be understood to be the test team) and need the tools and knowledge to work the system workflow [8].

The services don't have the service interface in unit testing [8], which this makes quality assurance team will have to be the implemented skills with to be able to produce good test, test objects, and the required test data. Another the service unit testing differences in input\output types with component testing and that complex [6], more test data generation techniques work on simple data, but in reality in complex systems such as SOA systems have complex inputs such as XML also used to inputs.

## 2.2. Non-Functional Testing Challenges

Non-functional properties of the system such as: availability, performance, applicability, maintenance capabilities, and portability. SOA has problems such as non-functional testing is impossible or difficult to determine a service workload parameters at service level agreement (SLA), the existing problems in the network and impact on system performance, system security (due to the decentralized system and the system is in the distributed system with different frameworks) [8]. The challenges of the test reliability can be reported [9]:

1- It may need a lot of time to test in the real execution as the faults may not occur in many situations.
2- Web services may involve many outside service providers who charge their service provided and it may a lot of cost.
3- The services in BPEL from different organizations should cooperate to achieve business goals. Their execution should be transactional.
4- Traditional faults generation techniques can only generate low level error regardless the business process, so that it makes difficult to test the whole process because the faults generated may not affect the business process intentionally.

## 2.3. Existing Solutions

The solutions of the functional testing challenges can be reported:

Among the proposed solutions for problems and challenges in the research include the functional testing procedures are updated at SOA Systems, and the existing methods are automated. The tools is used to performing complex actions and the integrity of system is able to management are produced. The monitoring system operated at all levels, and used the ESB capabilities for functional testing system. If needed and usability of other existing technologies (such as JMS middleware) was used.

Another solution is to improve the system new implementations [10]. Futures of the Internet need to make SOA testbed for large systems and reduce their cost of test. The testbed be required to validate and the integrity of the future Internet. The network should be loosely coupled, networks will be heterogeneous, and there will have capabilities in the program layer, and complexity will be greater in the composition. The systems need to automatic composition of services and management processes, there need to combine the testbed capabilities.

At the beginning of creating a production SOA system, the test team haven't high collaboration, but must be present to understand the goals and business processes, in order to properly and efficiently carry out the test, and that prevent the additional costs to the system development process.

## 2.4. Summarized

Table 1 shows the SOA Testing challenges with testing levels in the SOA systems.

Table 1. SOA Testing challenges with testing levels of the SOA systems.

| Challenges | Testing level | | | | |
| --- | --- | --- | --- | --- | --- |
| | Functional testing | Unit testing | Integration testing | Regression testing | Non-functional testing |
| dynamicity and adaptiveness | Parallel testing is hard, test automation | - | Integration will be in the run of system and Deploy and service description testing is hard | Regression testing is needed, test automation | Security of system will be low, test automation |
| lack of observability of the service code and structure | Stakeholders Use of different platforms | White box testing unable in service | The testability limits and information is low of services | The determinate of test time and sections is needed test are hard | Reliability will be low and the create of parameters QoS is hard |
| lack of control | Rapid development and complexity of system | - | Distributed of system and Integration will be in the run of system | The determinate of test time and sections is needed test are hard, Regression testing is needed | Efficiency, security, and reliability will be low |
| lack of trust | The change of system will be in run and complexity is high | - | The test will be in the run of system services | The change of system | Parameters QoS is offered, reliability challenges |
| new aspects of testing | The create of new approach and new tools | The create of new approach and new tools | The create of new approach and new tools | The create of new approach and new tools | The create of new approach and new tools |
| test cost | The test cost with traditional testing is high | The test cost with traditional testing is high | The test cost with traditional testing is high | The test cost with traditional testing is high | The test cost with traditional testing is high |
| different stakeholders | Software platforms, Distributed of system | - | The determinate of change is hard | The determinate of change time is hard | Security and reliability will be low |

## 3. SOA SYSTEMS TESTING

SOA system testing should be performed of aspects functional testing and non-functional testing, the functional testing has different levels, the levels include unit testing in individual services and the combined services, integration testing and regression testing.

This part of paper is checked SOA system testing of the functional aspects (in levels) and non-functional aspects, and then the test automation tools is described in the SOA system. Then automatic test data generation techniques are reviewed for traditional software. Finally, these results are shown in tables.

### 3.1 Unit Testing

Numbers of investigations in the unit testing have been active for test automation, in their attempted to automate process or processes of testing. The researches [11, 12, 13, 14, 15, 16] performed unit testing on WSDL file.

### 3.2 Integration Testing

Numbers of existing researches [17, 18, 19, 20, 21, 22, 23] performed BPEL-based testing in the system with operations graph.

In studies [24 and 25] implemented test at combining web services used to high level Petri net and specifications BPEL. You can generate test cases from web service automata (WSA) automatically [26]; WSA can be used to define the operational logic in BPEL.
Numbers of test frameworks have been prepared for SOA testing, than these are performed the SOA testing with the best way [27, 28].

The DFTT4CWS tools automatically find unusual data flow [29] and the test paths is generated data flow testing with all of cover criteria types. In reference [30], BPEL file is mapped DOM object tree to the EMF activities tree. The WebMov is set of tools modelling, evaluates and tests web services composition [31]. One paper was expressed computational strategy for the generation complete computational paths of BPEL based on Tabu Search and Genetic Algorithms to generate test data [32].

The research [33] provided an approach to design test cases based on functional properties of high-level business process model. The study [34] proposed an approach for reducing the costs to test such applications, and how can semantic stubs enable the client test suite to be partitioned into subsets, some of which don't use to execute remote services. Model driven approach is presented in [35], this approach to generate executable test cases from the given express business processes.

### 3.3 Regression Testing

Researches [36, 37] are proposed an approach to determine the changes use to extensible BPEL flow graph (XBFG) of control flow and to compare the paths in a new version of service composition with the old version.

Testing rules and monitoring rules include: checking the functional characteristics services, checking quality of service (QoS), checking interoperability services and service evolution [38].

### 3.4 Non-Functional Testing

Non-functional requirements include [39]: the needed data to fulfill the monitoring goal is intercepted. Monitoring mechanisms attempt the performance isn't influenced of unmonitored elements and performance is influenced of the monitored elements remains to be minimal. The changes responses are in the monitoring goal and environment topology. Instrumentation must be transparent and performed on demand.

System security is one of the characteristics non-functional SOA systems. The paper [40] presented a preliminary approach towards an evaluation framework for SOA security testing tools.

A research proposes a technique on how do reliability test define of composite service in BPEL from the view of business semantics with little cost using fault injection [9]. This paper focus on how the reliability problems find relate with business process, the called semantics as the problems are not pure coding error but faults related to business process. In addition, the behaviour of composite services in BPEL is analyzed when there are faults in the orchestrated services invoked.

### 3.5 Automatic testing tools

Numbers of existing produced tools was created to test SOA systems automatically. For example TASSA is a framework for automatic testing in functional and non-functional specifications of service-based applications [41]. It provides end-to-end testing of Service layer, Service Composition and coordination and business process. Another tool is WSOTF presented for the automated testing [42]. WSOTF is an automatic conformance testing tool with timing constraints from a formal specification of web services composition that is implemented by an online testing algorithm.

In the study [43] is expressed test approach described in BPEL web service composition. The paper [44] is proposed to generate a testbed for service-oriented systems that takes into account a mobility model of nodes in the network which the accessed services are deployed.

The study [45] is a framework and its supporting tool for automatically generating and executing web-service requests and analyzing the subsequent request-response pairs.

The study [46] is proposed an approach to combine the accessibility technologies in graphical applications (GAPs) for a visualization mechanism enables nonprogrammers to generate unit test cases in web services by drag-and-drop operations on graphical user interface (GUI). In the reference [47] is testing techniques to generate a set of test cases for web services automatically. The techniques presented here explore data perturbation of Web services messages upon data types, integrity and consistency.

### 3.6 Summarized

In the paper [48] is expressed a survey to explore cloud services testing methods. The paper [49] is expressed a review to identify SOA testing researches with dynamic binding, that paper performed manually and automatically search in journals, conferences and etc.

Methods described in Table 2 with different levels of test coverage, SOA system testing and see in them. In A service may be provided in the composition of services using BPEL file, the reason test on the some parts of this table is on the integration testing (the BPEL file) also put on the unit testing capabilities.

Table 2. SOA systems testing Methods at levels testing

| Method | Level Testing | | | |
| --- | --- | --- | --- | --- |
| | Unit Testing | Integration Testing | Regression Testing | Non-Functional Testing |
| DOM tree [11, 14], DbC [13, 16], WS-TAXI [15], GAP [46], GenAutoWS [47], [45] | Generate test cases by WSDL | - | - | - |
| SAT Solver [21] | - | Generate test cases by processes and user activity | Use to save test cases | - |
| CPM [33], [20, 27, 28, 29, 34, 35] | - | Generate test cases for processes business | - | - |
| SXM [22, 23], BTA [25], TestGen-IF [43], WSA [26], [17, 18, 19, 24, 30], Tabu [32] | BPEL testing | Generate test cases for BPEL service | - | - |
| TASSA [41] | Layer service testing | Generate test cases at orchestration and BPEL service | - | Layers testing, Coordination and service composition |
| WSOTF [42] | Analysis WSDL | Generate test cases at specification system | - | - |
| XBFG [36,37] | - | - | Select test cases | - |
| [40] | - | - | - | Security |
| [9] | - | - | - | Reliability |
| [44] | - | - | - | Generate specification in mobile system model |

## 4. CONCLUSIONS

The part of tow this paper was expressed challenges and limitations of the SOA systems testing, and the existing solutions to solve some of them. The SOA systems testing challenges include: dynamicity and adaptiveness, lack of observability of the service code and structure, lack of control, lack of trust, new aspects of testing, test cost, different stakeholders. The part of three this paper was expressed SOA systems testing at two aspects of functional testing and non-functional testing. The functional testing was investigated in three different parts of the unit, integration and regression. In the last of chapters the results of these researches showed in the relevant tables.

International Journal of Software Engineering & Applications (IJSEA), Vol.3, No.6, November 2012

Manually test cases generation and manually test operations are a difficult and time consuming and the dynamic nature of the SOA systems cause the test cases is generated after some time lose their usability. To resolve this problem need to create a dynamic and automatic way to generate test cases in orchestration of SOA. Future works will propose specific approaches for specific software in the software logic or improve existing approaches for specific software. Another create test cases generation framework for SOA systems. Future work can be integration of available tools.

## ACKNOWLEDGEMENTS

The author would like to thank specially Dr. Seyed Hasan Mirian Hossienabadi who has extended his support for successful completion of this paper.